\documentclass[twocolumn,secnumarabic,amssymb, nobibnotes, aps, prd,longbibliography]{revtex4-2}

\usepackage{latexsym}
\usepackage{graphicx}
\usepackage{amsmath}
\usepackage{amssymb}
\usepackage{amsfonts}
\usepackage{times}
\usepackage{xspace} 
\usepackage[usenames]{color}
\usepackage{dcolumn}
\usepackage{bm}
\usepackage{mathrsfs}
\usepackage{tikz}
\usepackage{appendix}
\usepackage{breqn}
\usepackage[]{units}       % for units environment

\usepackage{subfigure}		% for subfigures
\usepackage[normalem]{ulem} % for strikeout

\usepackage{hyperref} % for hyperlinks

\usepackage[tracking=true]{microtype}
\SetTracking{}{500}
\SetTracking{encoding={*}, shape=sc}{40}
\AtBeginDocument{%
    \newwrite\bibnotes
    \def\bibnotesext{notesNotes.bib}
    \immediate\openout\bibnotes=\jobname\bibnotesext
    \immediate\write\bibnotes{@CONTROL{REVTEX41Control}}
    \immediate\write\bibnotes{@CONTROL{%
    apsrev41Control,author="08",editor="1",pages="1",title="0",year="1"}}
     \if@filesw
     \immediate\write\@auxout{\string\citation{apsrev41Control}}%
    \fi
}%
\makeatletter
\let\cat@comma@active\@empty
\makeatother

\begin{document} 
 
\title{Gravitational wave signatures from dark sector interactions}

\author{Reginald Christian Bernardo}
\affiliation{National Institute of Physics, University of the Philippines, Diliman, Quezon City 1101, Philippines}
\email{rbernardo@nip.upd.edu.ph}
\date{\today}

\begin{abstract}
We show that the gravitational waves generated by the perturbations of general relativistic black holes can be considered as a direct probe of the existence of dark sector interactions. Working within the framework of Horndeski theory and linear perturbations, we show that dark sector interactions effectively reduce to an interaction charge that influences both scalar and tensor waveforms. Furthermore, we show that the total dark matter field, including the effects of dark sector interactions, satisfies a conservation equation embodying the equivalence principle. We exploit this realization to setup the Regge-Wheeler equation and the coupled Zerilli and scalar wave equations for a Schwarzschild-(anti) de Sitter black hole. We then present numerical integration of the coupled even-parity wave equations for the case of a dark matter particle falling straight down into a Schwarzschild black hole.
\end{abstract}

\maketitle

%--------------------------------------------------
% the main text of your paper begins here
%--------------------------------------------------

\section{Introduction}
\label{sec:introduction}

The general theory of relativity (GR) is by far the best theory of the observable Universe. However, it requires the addition of dark, a.k.a., optically invisible, fluids at cosmological distances to explain the observed late-time cosmic acceleration and the preceeding matter era. The concordance $\Lambda$CDM model, in which dark energy is treated as a cosmological constant, also fails to give a definitive fundamental description of the dark fields and sidesteps the coincidence and cosmological constant problems. The recent recognition of the tensions between the early and late measurements of the Hubble parameter and matter power spectrum has put the $\Lambda$CDM model in an even tighter spot \cite{local_hubble_riess, early_hubble_planck, hubble_tension_verde}. To properly assess the regime of validity of the $\Lambda$CDM model, it makes sense to pay attention to what else can the dynamical theory of the Universe be or rather consider alternative theories of gravity.

The simplest, and arguably most compelling, alternative theories are known collectively as scalar-tensor theories \cite{alternative_gravity_clifton, alternative_gravity_joyce, alternative_gravity_koyama, review_dark_energy_bamba}. In this class, the scalar field stands as a surrogate to dark energy, endowing it with dynamics, and so brings an additional propagating degree of freedom (d.o.f.) in addition to the two tensor polarizations carried by the metric. However, arbitrary couplings of the scalar and tensor fields in the action generally come with gravitational waves (GWs) travelling at a speed different than that of light. The observation of GWs from the binary neutron star merger event, GW170817, and its optical counterpart has constrained this speed difference to one part in $10^{15}$ \cite{gw_170817_ligo}. Viable alternative theories of gravity should then have GWs propagating at the speed of light with little wiggle room for error \cite{lombriser2016breaking, lombriser2017challenges, dark_energy_creminelli, dark_energy_ezquiaga, st_horndeski_cosmology_baker, st_horndeski_cosmology_sakstein, st_dark_energy_tsujikawa, horndeski_review_kobayashi2019, st_horndeski_copeland2018, ggc_dark_energy_kase, ferreira2019cosmological}. Notably, the existing parameter space of scalar-tensor theories remains to be quite large even with such restriction and continues to be interesting from a cosmological perspective. Furthermore, interest in dark sector interactions, coupling the dark matter and dark energy fields, has been renewed recently as a potential answer to the Hubble tension \cite{coupled_quint_pettorino, idm_richarte_1, idm_richarte_2, idm_richarte_3, idm_richarte_4, interacting_de_kase_tsujikawa, ide_proca_nakamura, scaling_kgb_frusciante, ide_dasilva, ide_fake_valentino, ide_closed_valentino, ide_pan, ide_tao_yang, ide_johnson_2, ide_gomez, ide_asghari, ide_kang, ide_yang, ide_jimenez, ide_marttens, ide_benetti, ide_bonilla, ide_johnson, ide_proca_gomez, realm_hubble_valentino}. The search for observational signatures that distinguish between theories with and without dark sector interactions is therefore of considerable importance in order to constrain alternative theories.

Black holes are powerful astrophysical laboratories for observing new d.o.f.s in action and discriminating between alternative theories of gravity. For one, these are the places where gravity is at its strongest and so probe the highly nonlinear and dynamical regime of a gravity theory. Also, the black holes in GR can be simply described by their mass, spin, and electric charge parameters, or ``hair'' \cite{no_hair_classic_bekenstein, no_hair_classic_israel, no_hair_classic_israel_2, no_hair_carter}. Non-GR d.o.f.s therefore generally allow a black hole to depend on additional parameters. Ironically, general relativistic black holes, i.e., those without any hair, appear to be generally permitted in alternative theories of gravity while hairy black holes seem to lack cosmological significance or suffer from strong coupling issues and other pathologies \cite{st_bd_hawking, st_no_hair_theorem_hui, st_no_hair_theorem_sotiriou_1, st_no_hair_theorem_sotiriou_2, st_black_holes_sotiriou, hairy_bhs_tfl, st_bh_pt_de_rham, no_hair_creminelli}. Nonetheless, general relativistic black holes continue to be interesting in alternative theories because their quasinormal spectra can reveal the existence of new d.o.f.s through a mixing with the tensor modes \cite{st_horndeski_qnm_tattersall, bh_spectro_tattersall, anomalous_qnms_lagos, qnms_conformal_bhs_chen, qnms_lessons_chen}. In scalar-tensor theories, this can be achieved via a conformal coupling of the scalar field with the Ricci scalar in the gravitational action in the physical frame. In this paper, we work with general relativistic black holes and consider it as a setting to directly probe interactions between dark matter and dark energy.

Dark sector interactions, by their very nature, can be observed indirectly by their influence on astronomical observations of the expansion history, cosmic microwave background, and large scale structure \cite{gw_ide_caprini, gw_ide_bachega, gw_ide_yang, gw_ide_yang_2}. On the other hand, even dark fields respond gravitationally and produce GWs whenever they interact. GWs therefore are a direct means of detecting dark sector interactions. Given the potentially legitimate role played by dark sector interactions in cosmology, it is perfectly valid to ask whether they can be probed with GW signals and in what setting. We consider this question in this paper and prepare the formalism for the perturbations of general relativistic black holes with dark sector interactions. We work with the most general Horndeski theory consistent with GW170817 (Eq. \eqref{eq:horndeski}) and accommodate two fairly general interaction Lagrangians (Eqs. \eqref{eq:Lc1} and \eqref{eq:Lc2}) that encompass most of the specific choices of dark sector interactions considered in the literature.

The rest of this work proceeds as follows. First, we present the gravitational action of Horndeski theory and the interaction Lagrangians which will represent dark sector interactions (Section \ref{subsec:action}). The covariant field equations are then presented (Section \ref{subsec:covariant_field_equations}) and shortly followed by a discussion of its cosmological limit (Section \ref{subsec:cosmology}) and general relativistic black hole solutions (Section \ref{subsec:black_holes}). In Section \ref{sec:bh_perturbations}, we discuss general relativistic black hole perturbations and obtain the main results of this paper. We obtain linearized field equations in covariant form (Section \ref{subsec:linearized_equations}) and its spherical harmonics analysis in Schwarzschild-(anti) de Sitter black hole leading to the sourced Regge-Wheeler equation and coupled Zerilli and scalar wave equations. To make the results explicit, we integrate the coupled Zerilli and scalar wave equations for a dark matter particle falling straight down towards a Schwarzschild black hole (Section \ref{sec:scalar_tensor_waveforms}). We conclude with a discussion of the limitations of this analysis and future work (Section \ref{sec:conclusions}). In Appendices \ref{sec:set_plunge} and \ref{sec:integration}, we provide the stress-energy tensor for a radially-falling particle and the recipe used for the numerical integration in this paper.

\textit{Conventions}. We work with the mostly-plus metric signature, $\left(-, +, +, +\right)$, and in geometrized units, $c = 8 \pi G = 1$. Spacetime (two-sphere) indices are denoted by lowercase (uppercase) latin symbols. For brevity, we denote the covariant derivatives of a spacetime scalar $\Phi$ as $\Phi_a$, e.g., $\nabla_a \Phi = \Phi_a, \nabla_{a} \nabla_{b} \Phi = \Phi_{ab}$. Sum and integral over the spherical harmonic modes $\left(l,m\right)$ and frequencies $\omega$ are implicit. The reader interested in all of the mathematical details of this paper is encouraged to download the supplementary Mathematica notebooks in the \href{https://github.com/reggiebernardo/notebooks}{author's github repository}.

\section{Horndeski theory}
\label{sec:horndeski_theory}

In this section, we present Horndeski theory with dark sector interactions and discuss its observational constraints. The covariant field equations, its cosmological limit, and general relativistic black hole solutions are also briefly discussed.

\subsection{Action}
\label{subsec:action}

We consider the sector of Horndeski theory consistent with the observation GW170817. This is given by the gravitational action \cite{dark_energy_creminelli, dark_energy_ezquiaga, st_dark_energy_tsujikawa, horndeski_review_kobayashi2019, ggc_dark_energy_kase, ferreira2019cosmological}
\begin{equation}
\label{eq:horndeski}
S_g = \int d^4 x \sqrt{-g} \left[ F\left(\phi\right) R + V \left( \phi, X \right) - G \left( \phi , X \right) \Box \phi \right]
\end{equation}
where $g_{ab}$ is the metric, $R$ is the Ricci scalar, and $F$, $V$, and $G$ are arbitrary potentials depending on the scalar field $\phi$ and its kinetic density $X = -\left(\partial\phi\right)^2 /2$. We refer to $F$, $V$, and $G$ as the conformal, $k$-essence, and braiding potentials, respectively. The above action can also be transformed to the so-called kinetic gravity braiding class of theories when one goes to the Einstein frame \cite{st_galileon_inflation_deffayet}. In this conformally equivalent frame, the tensor modes also propagate on the light cone and the matter fields become nonminimally-coupled to the conformal metric. For this work, we restrict our attention to the physical frame (Eq. \eqref{eq:horndeski}) where the metric and scalar field are $g_{ab}$ and $\phi$, respectively.

In addition to Eq. \eqref{eq:horndeski}, we consider the \textit{free} dark matter sector to be given by the Schutz-Sorkin action \cite{stability_matter_defelice}
\begin{equation}
\label{eq:schutz_sorkin}
S_{\text{dm}} = - \int d^4 x \left( \sqrt{-g} \rho \left( n \right) + J^b \nabla_b l \right) 
\end{equation}
where $\rho$, $n$, and $J^a$ are the CDM's energy density, particle number, and particle current and $l$ is a Lagrange multiplier. The CDM's fluid variables satisfy the following relations:
\begin{eqnarray}
\label{eq:ndef} n &=& \sqrt{ J_a J^a / g } \\
\label{eq:udef} u^b &=& J^b/ \left( n \sqrt{ -g } \right) \\
\label{eq:pdef} P &=& n \left( \partial \rho / \partial n \right) - \rho \left( n \right)
\end{eqnarray}
where $u^b$ and $P$ are the CDM's velocity and pressure, respectively. Note that these relations guarantee $u_b u^b = -1$. It must be noted that other Lagrangian representations of the dark matter sector exist \cite{PhysRevLett.93.011301, doi:10.1142/S0217732307025893, PhysRevD.89.064059, Gleyzes_2015} and that the Schutz-Sorkin action is a preferential choice. Nonetheless, Eq. \eqref{eq:schutz_sorkin} has been proven to be useful in the analyses of cosmological perturbations in modified gravity \cite{stability_matter_defelice, ggc_dark_energy_kase, scaling_kgb_frusciante, st_dark_energy_tsujikawa, interacting_de_kase_tsujikawa} and is a suitable companion to the interactions to be discussed. We add to this the general interaction Lagrangians given by \cite{interacting_de_kase_tsujikawa}
\begin{equation}
\label{eq:Lc1}
\mathcal{L}_1 = - I_1 \left( \phi, X \right) \rho \left( n \right)
\end{equation}
and
\begin{equation}
\label{eq:Lc2}
\mathcal{L}_2 = - I_2 \left( n ,\phi,  X \right) J^a \partial_a \phi / \sqrt{-g} 
\end{equation}
where $I_1$ and $I_2$ are two arbitrary interaction potentials coupling the dark sector. This representation of the dark sector interaction encompasses most of the particular forms of interacting dark energy studied in the literature \cite{idm_richarte_4, ide_yang, ide_benetti, ide_kang, ide_gomez, ide_fake_valentino, ide_pan}. For example, in Refs. \cite{coupled_quint_pettorino, ide_gomez}, the interaction considered lead to a heat term $\mathcal{Q} = -\beta \rho_{\text{dm}} \dot{\phi}$ where $\beta$ is a constant. This can be merely recognized as special case of the first term of the heat term (Eq. \eqref{eq:heat_term_cosmo}) arising from the interaction potentials given by Eqs. \eqref{eq:Lc1} and \eqref{eq:Lc2}.

Any visible matter can always be considered by including an action $S_{\text{vm}}$ with a stress-energy tensor (SET) given by
\begin{equation}
T^{(\text{vm})}_{ab} = - \left( 2 / \sqrt{-g} \right) \left( \delta S_{\text{vm}} / \delta g^{ab} \right) .
\end{equation}
The variation of the total action with respect to the fields and multipliers straightforwardly gives the field equations of the theory.

The theory \eqref{eq:horndeski} has tensor modes propagating on the light cone and evades the GW astronomy constraint on the GW speed. On the cosmology side, some of its sectors have been analyzed in detail with observational data. The Galileon ghost condensate \cite{peirone2019cosmological} and the generalized cubic covariant Galileon \cite{gccg_frusciante} appear to be as cosmologically viable as $\Lambda$CDM even with large cosmological datasets that have ruled out the cubic Galileon \cite{st_horndeski_galileon_barreira_1, st_horndeski_galileon_barreira_2, st_horndeski_renk, st_horndeski_galileon_peirone}. Interestingly, special observational limits of cosmologically-modified gravity have been singled out in Refs. \cite{no_slip_linder, no_slip_cmb_brush, no_run_linder, limited_modified_gravity}. General cosmological constraints on the theory \eqref{eq:horndeski} also exist through parametrizations of the matter/light linear gravitational potentials \cite{10.1093/mnras/sts493, early_hubble_planck} and through effective field theory routes \cite{horndeski_constraint_noller2018, horndeski_constraint_komatsu2019}. A worthy alternative theory should also be compatible with Solar system tests and therefore must keep a screening mechanism \cite{screening_schmidt, screening_mcmanus, st_horndeski_vainshtein_dima, kgb_vainshtein_anson}.

\subsection{Covariant field equations}
\label{subsec:covariant_field_equations}

The modified Einstein equation is given by
\begin{equation}
\label{eq:einstein_eq}
E_{bc} = \dfrac{T^{(\text{vm})}_{bc}}{2}
\end{equation}
where
\begin{equation}
\begin{split}
E_{bc } = F G_{bc} & + F' \left( g_{bc} \Box \phi - \phi_{bc} \right) + F'' \left( g_{bc} \Box \phi - \phi_b \phi_c \right) \\
& - \dfrac{V}{2} g_{bc} - \dfrac{V_X}{2} \phi_b \phi_c + G_\phi \left( \phi_b \phi_c + X g_{bc} \right) \\
& + \dfrac{G_X}{2} \left( \phi_b \phi_c \Box \phi + \phi_{bc} \phi_{ed} \phi^e \phi^d - 2 \phi_{(b} \phi_{c)d} \phi^d \right) \\
& + \dfrac{ \tilde{ \rho } }{2} \dfrac{I_{1X}}{1 + I_1} \phi_b \phi_c - \dfrac{n}{2} I_{2X} u^d \phi_d \phi_b \phi_c - \dfrac{\tilde{T}^{(\text{dm})}_{bc}}{2} ,
\end{split}
\end{equation}
\begin{equation}
\label{eq:T_tilde}
\tilde{T}^{(\text{dm})}_{bc} = \left( \tilde{\rho} + \tilde{P} \right) u_b u_c + \tilde{P} g_{bc} ,
\end{equation}
\begin{equation}
\label{eq:rho_tilde}
\tilde{\rho} = \left( 1 + I_1 \right) \rho ,
\end{equation}
and
\begin{equation}
\label{eq:p_tilde}
\tilde{P} = \left( 1 + I_1 \right) P - n^2 u^d \phi_d I_{2n} .
\end{equation}
The subscripts appearing in the potentials $V, G, I_j$ denote explicit derivatives with respect to their arguments, e.g., $V_X = \partial V/ \partial X$ and $I_{1X} = \partial I_1 / \partial X$. Primes on the conformal coupling $F$ denote derivatives with respect to $\phi$, e.g., $F' = d F /d\phi$ and $F'' = d^2 F/d \phi^2$. In the dark matter fields (Eqs. \eqref{eq:T_tilde}, \eqref{eq:rho_tilde}, and \eqref{eq:p_tilde}), we refer to the terms independent of and attached to the interaction potentials as the \textit{free} and \textit{interacting} parts, respectively, e.g., $\rho$, $I_1 \rho$, and $\tilde{\rho}$ are the free, interacting, and total dark matter energy density, respectively. In particular, a dark matter quantity $\tilde{\chi}$, with a tilde on top, e.g., $\tilde{\rho}$, $\tilde{P}$, and $\tilde{T}_{bc}^{(\text{dm})}$, refers to the total field, or rather the sum of its free and interacting components.

The scalar field equation is given by
\begin{equation}
\label{eq:sfe}
\mathcal{S}_\phi = 0
\end{equation}
where
\begin{equation}
\begin{split}
\mathcal{S}_\phi = V_X \Box \phi & + V_{XX} \phi_{ab}\phi^a\phi^b + V_\phi - 2X V_{\phi X} \\
& + G_X \left( \phi_{ab} \phi^{ab} + R_{bc} \phi^b\phi^c - \left( \Box \phi \right)^2 \right) \\
& + G_{XX} \phi^b \phi^c \left( \phi_{bc} \Box \phi - \phi_{dc} \phi_b^{\ d}  \right) \\
& - 2 \Box \phi G_\phi + 2 X G_{\phi \phi} \\
& + G_{\phi X}\phi^b \left( 2 \phi_{bc} \phi^c - \phi_b \Box \phi \right) + F' R \\
& + \tilde{\rho} \dfrac{I_{1XX}}{1 + I_1} \phi_{ab} \phi^a \phi^b - \tilde{\rho} \dfrac{I_{1\phi}}{1 + I_1} +2 X \tilde{\rho} \dfrac{I_{1 \phi X}}{1 + I_1} \\
& + I_{2X} \bigg( n \phi_b \phi_c \nabla^c u^b + u^b \phi_b \nabla_c \left( n \phi^c \right) \\
& \phantom{gggggggggggggggig} + 2n\phi_{bc} u^b \phi^c \bigg) \\
& - n u^b \phi_b \phi_{cd} \phi^c \phi^d I_{2XX} + n u^b \phi_b \phi_c \phi^c I_{2 \phi X} \\
& + n^2 \nabla_b u^b I_{2n} + n u^b \phi_b \phi^c n_c I_{2 n X} \\
& - \dfrac{I_{1X}}{1 + I_1} \bigg( \tilde{\rho} \Box \phi + \left( \dfrac{ \tilde{\rho} + \tilde{P} }{n} \right) \phi^b n_b \\
& \phantom{ggggggggggggg} + n I_{2n} u^d \phi_d \phi^c n_c \bigg) . 
\end{split}
\end{equation}
We identify the dark sector's stress-energy tensor, $T_{ab}^{(\text{D})}$, as
\begin{equation}
T_{ab}^{(\text{D})} = 2 \left( F G_{ab} - E_{ab} \right) 
\end{equation} 
so that the modified Einstein equation (Eq. \eqref{eq:einstein_eq}) can be written as
\begin{equation}
F G_{bc} - \dfrac{T_{ab}^{(\text{D})}}{2} = \dfrac{T^{(\text{vm})}_{bc}}{2} .
\end{equation}
The dark matter's field equation will be determined by the Bianchi identity, $\nabla_b G^{ab} = 0$, and the scalar field equation.

\subsection{Cosmology}
\label{subsec:cosmology}

On cosmological distances, assuming a spatially-flat Friedmann-Robertson-Walker metric,
\begin{equation}
\label{eq:frw}
ds^2 = - dt^2 + a\left( t \right)^2 d \vec{x}^2 ,
\end{equation}
and a comoving scalar, $\phi = \phi\left(t\right)$, it can be shown that the dark energy and dark matter fluids satisfy 
\begin{equation}
\label{eq:de_eq_cosmo}
\dot{ \varrho }_\phi + 3 H \left( t \right) \left( \varrho_\phi + \mathcal{P}_\phi \right) = - \mathcal{Q} 
\end{equation}
and
\begin{equation}
\label{eq:cdm_eq_cosmo}
\dot{ \tilde{\rho} } + 3 H \left(  t \right) \left( \tilde{\rho} + \tilde{P} \right) = \mathcal{Q} ,
\end{equation}
respectively, where an overdot denotes differentiation with respect to the cosmic time $t$ and $H\left(t\right) = \dot{a}/a$ is the Hubble parameter. In Eqs. \eqref{eq:de_eq_cosmo} and \eqref{eq:cdm_eq_cosmo}, the energy densities and pressures are given by \cite{interacting_de_kase_tsujikawa}
\begin{equation}
\begin{split}
\varrho_\phi = & - V + \dot{\phi}^2 V_X + 3 H \dot{\phi}^3 G_X - \dot{\phi}^2 G_\phi \\
& + 3 \left(1 - 2F \right)H^2 - 6 H F' \dot{\phi} \\
& - \tilde{\rho} \dot{\phi}^2 \dfrac{I_{1X}}{ 1 + I_1 } + n \dot{\phi}^3 I_{2X} ,
\end{split}
\end{equation}
\begin{equation}
\begin{split}
\mathcal{P}_\phi = V &- \dot{\phi}^2 \ddot{\phi} G_X - \dot{\phi}^2 G_\phi - H^2 \left( 1 - 2 F \right) \\
& + 4 H \dot{\phi} F' - \dfrac{\ddot{a}}{a} \left( 1 - 2F \right) + 2\ddot{\phi} F' + 2 \dot{\phi}^2 F'' ,
\end{split}
\end{equation}
\begin{equation}
\tilde{\rho} = \left( 1 + I_1 \right) \rho ,
\end{equation}
\begin{equation}
\tilde{P} = \left( 1 + I_1 \right) P - n^2 \dot{\phi} I_{2n} ,
\end{equation}
and the heat term is given by
\begin{equation}
\label{eq:heat_term_cosmo}
\mathcal{Q} = \tilde{\rho} \dot{\phi} \dfrac{I_{1\phi}}{1 + I_1} + \tilde{\rho} \dot{\phi} \ddot{\phi} \dfrac{I_{1X}}{1 + I_1} - 3 n^2 H \dot{\phi} I_{2n} .  
\end{equation}

It is important to point out that no particular form of the interaction potentials have been assumed in Eqs. \eqref{eq:de_eq_cosmo} and \eqref{eq:cdm_eq_cosmo}. The heat term (Eq. \eqref{eq:heat_term_cosmo}) can take on any form and encompasses most of the specific forms assumed in the literature \cite{idm_richarte_4, ide_yang, ide_benetti, ide_kang, ide_gomez, ide_fake_valentino, ide_pan}.

\subsection{Black holes}
\label{subsec:black_holes}

We consider bald black holes, i.e., general relativistic vacuum solution
\begin{equation}
\label{eq:gr_bh}
G_{ab} + \Lambda g_{ab} = 0
\end{equation}
with trivial scalar and dark matter fields,
\begin{equation}
\label{eq:trivial_scalar}
\phi \left( x \right) = \varphi ,
\end{equation}
\begin{equation}
\label{eq:trivial_cdm}
\tilde{\rho} , \tilde{P}, n = 0 ,
\end{equation}
where $\varphi$ is a constant. It is easy to show that this is an exact solution to the field equations provided that
\begin{equation}
V = - 2 \Lambda F
\end{equation}
and
\begin{equation}
V_\phi = - 4 \Lambda F' .
\end{equation}

This is a good place to comment on the distinction between general relativistic black holes and stealth black holes in the literature \cite{st_horndeski_solutions_babichev_0, st_horndeski_qnm_tattersall}. Both solutions are described by a metric that satisfies Eq. \eqref{eq:gr_bh}. However, stealth black holes come with nontrivial fields or hair; on the other hand, general relativistic black holes, which may also be referred to as bald black holes, are accompanied by constant or trivial fields (e.g., Eqs. \eqref{eq:trivial_scalar} and \eqref{eq:trivial_cdm}).

In an astrophysical setting, a black hole may be embraced by a nearly constant scalar field and is generally expected to accrete dark matter from its surroundings. The general relativistic black hole can be regarded as an idealized first approximation to this realistic scenario that caters to a simple, analytical solution, allowing various physical analysis, e.g., quasinormal spectra \cite{st_horndeski_qnm_tattersall, bh_spectro_tattersall, anomalous_qnms_lagos}. The existing formalisms for treating dark matter accretion into black holes also support the continuity of the solution in going from zero to small densities \cite{dm_accretion_mach, dm_accretion_bamber}.

The general stationary solution to Eq. \eqref{eq:gr_bh} is given by the Kerr-(anti) de Sitter solution \cite{st_kerr_bhs_charmousis, stealth_bernardo}. The discussion of Section \ref{subsec:linearized_equations} should then be understood within this context. Only in the derivation of the Regge-Wheeler, Zerilli, and scalar wave equations in Sections \ref{subsec:odd_parity} and \ref{subsec:even_parity} will the focus be restricted to nonrotating black holes. In this case, we consider the static and spherically symmetric geometry given by
\begin{equation}
\label{eq:black_hole_metric}
ds^2 = - B\left( r\right) dt^2 + \dfrac{dr^2}{B\left(r\right)} + r^2 \left( d\theta^2 + \sin^2 \theta d \alpha^2 \right) .
\end{equation}
Eq. \eqref{eq:gr_bh} is satisfied by the Schwarzschild-(anti) de Sitter solution given by
\begin{equation}
B \left( r \right) = 1 - \dfrac{2M}{r} - \dfrac{ \Lambda r^2 }{3}
\end{equation}
where $M$ is an integration constant.

\section{Black hole perturbations}
\label{sec:bh_perturbations}

In this section, we obtain the main results of this paper. We obtain the linearized field equations on general relativistic black holes (Section \ref{subsec:linearized_equations}) and derive the odd- and even-parity master equations (Sections \ref{subsec:odd_parity} and \ref{subsec:even_parity}).

\subsection{Linearized equations}
\label{subsec:linearized_equations}

To study black hole perturbations, we substitute the perturbations
\begin{equation}
g_{ab} \rightarrow g_{ab} + h_{ab} ,
\end{equation}
\begin{equation}
\phi \rightarrow \phi + \psi ,
\end{equation}
\begin{equation}
\tilde{\rho} \rightarrow \tilde{\rho} + \delta \tilde{\rho} ,
\end{equation}
\begin{equation}
\tilde{P} \rightarrow \tilde{P} + \delta \tilde{P} ,
\end{equation}
\begin{equation}
n \rightarrow n + \delta n ,
\end{equation}
and
\begin{equation}
u^b \rightarrow u^b + \delta u^b 
\end{equation}
to the field equations and let the background satisfy the general relativistic black hole solution (Eqs. \eqref{eq:gr_bh}, \eqref{eq:trivial_scalar}, and \eqref{eq:trivial_cdm}). We remind that the general relativistic black hole comes with a constant scalar field and vanishing dark matter. In practice, this corresponds to first performing the perturbative expansion of the field equations for general fields, and then taking the zeroth order fields to satisfy Eqs. \eqref{eq:gr_bh}, \eqref{eq:trivial_scalar}, and \eqref{eq:trivial_cdm}. The reader interested in the full, background-agnostic, perturbative expressions may refer to the Mathematica notebook \textit{bald\_bh\_perturb.nb} in the \href{https://github.com/reggiebernardo/notebooks}{author's github repository}.

This straightforward but admitedly tedious task will lead to the following final expressions for the linearized evolution equations of the perturbations $\left( h_{ab}, \psi, \delta \tilde{\rho}, \delta \tilde{P} \right)$. The tensor perturbation $h_{ab}$ will be governed by
\begin{equation}
\label{eq:h_eq}
F \left( \delta G_{bc} + \Lambda h_{bc} \right) + \Lambda F' \psi g_{bc} + F' \psi_{bc} + F' \Box \psi g_{bc} = \dfrac{\delta \tilde{T}^{(\text{dm})}_{bc}}{2} 
\end{equation}
where
\begin{equation}
\label{eq:dm_perturb_set}
\delta \tilde{T}^{(\text{dm})}_{bc} = \left( \delta \tilde{ \rho } + \delta \tilde{ P } \right) u_b u_c + \delta \tilde{P} g_{bc} .
\end{equation}
This is the linearized modified Einstein equation when there is a conformal coupling to the Ricci scalar in the action. An important distinction should be made when interpreting this and comparing this with the noninteracting case: it is the \textit{total} dark matter field, or rather its fluid variables, that the metric perturbation responds to (recall Eqs. \eqref{eq:rho_tilde} and \eqref{eq:p_tilde}). On the other hand, the scalar perturbation $\psi$ will be governed by
\begin{equation}
\label{eq:psi_eq}
\Box \psi - \mu^2 \psi =  s \delta \tilde{T}^{(\text{dm})} + q \delta \tilde{ \rho }
\end{equation}
where the effective mass $\mu$, conformal coupling constant $s$, and interaction charge $q$ are given by
\begin{equation}
\label{eq:mu_def}
\mu^2 = \dfrac{ V_{\phi\phi} + 4 \Lambda F'' + 4 \Lambda \left( F^{\prime 2} / F \right) }{ 3 \left( F^{\prime 2} / F \right) + V_X - 2 G_\phi } ,
\end{equation}
\begin{equation}
\label{eq:s_def}
s = \dfrac{ \left(F^{\prime 2} / \left( 2 F \right) \right) }{ 3 \left( F^{\prime 2} / F \right) + V_X - 2 G_\phi } ,
\end{equation}
and
\begin{equation}
\label{eq:q_def}
q = \dfrac{I_{1\phi}}{1 + I_1} \left( 3 \left( F^{\prime 2} / F \right) + V_X - 2 G_\phi \right)^{-1} .
\end{equation}
The mass $\mu$ is a generalization of the expression obtained in Ref. \cite{st_horndeski_qnm_tattersall} to include nonzero $\Lambda$. The constant $s \sim F'$ is responsible for the conformal coupling between the scalar $\psi$ and tensor $h_{ab}$ modes. Specifically, when $s \neq 0$, then the tensor modes' quasinormal spectrum will be contaminated by the scalar and should be observable in the ringdown phase of black hole binaries \cite{st_horndeski_qnm_tattersall, bh_spectro_tattersall, anomalous_qnms_lagos}. Most importantly, we introduce the dark sector's interaction charge $q$ (Eq. \eqref{eq:q_def}) which additionally sources the scalar perturbation $\psi$. This quantity is the leftover of dark sector interactions on a general relativistic black hole background. Its effect on the waveform is the main interest of this paper.

The dark matter's evolution equation turns out to be
\begin{equation}
\label{eq:cdm_conservation}
\nabla^b \delta \tilde{T}^{(\text{dm})}_{bc} = 0 .
\end{equation}
This was derived starting from the Bianchi identity, $\nabla^b G_{ab} \sim \nabla^b \left( T^{(\text{D})}_{ab} / F\left(\phi\right) \right) = 0$, and then performing the perturbative expansion. It is noteworthy that Eq. \eqref{eq:cdm_conservation} is of the exact same form as the free dark matter field's conservation equation except that it is the total field, including both the free and interacting parts, that is considered in this equation. Recall that $\tilde{T}^{(\text{dm})}_{bc}$ denotes the dark matter's total stress-energy tensor (Eqs. \eqref{eq:T_tilde}, \eqref{eq:rho_tilde}, and \eqref{eq:p_tilde}). Its free and interacting components correspond to the terms independent of and proportional to the interaction potentials, respectively. This result (Eq. \eqref{eq:cdm_conservation}) can be considered as a manifestation of the equivalence principle for dark matter perturbations even when there are dark sector couplings. We can exploit this remarkable result in order to simulate the direct effect of dark sector interactions on the scalar and tensor waveforms. To get to this, we setup the master equation for the odd- and even-parity sectors of the black hole perturbations in the next two sections.

It is important to point out that the interaction charge does not depend on the potential $I_2$. The dependence on $I_2$ vanished because $I_2$ always comes with factors of $n$ or $\phi_b$ wherever it appeared in the field equations. This result means that dark sector interactions of the form of Eq. \eqref{eq:Lc2} cannot be probed using general relativistic black holes.

The main effect of the interaction charge $q$ can be fleshed out by considering the dark matter perturbation as a particle. In this case, $\delta \tilde{P} \ll \delta \tilde{\rho}$, and so, by virtue of Eq. \eqref{eq:cdm_conservation}, the dark matter particle will be on the geodesics of the black hole. Substituting the trace of Eq. \eqref{eq:dm_perturb_set} into Eq. \eqref{eq:psi_eq}, we obtain
\begin{equation}
\Box \psi - \mu^2 \psi = \left( q - s \right) \delta \tilde{\rho} + O \left( \delta \tilde{P} / \delta \tilde{\rho} \right) .
\end{equation}
The interpretation of this equation is clear. Dark sector interactions directly enhance or deteriorate the size of the scalar waveform $\psi$ by a factor $-q / s$
\footnote{
To get to this factor, simply compare the sizes of the wave with $q \neq 0$ and with $q = 0$, i.e., $ \left( q - s \right) - \left( - s \right) / \left( -s \right)  =  -q/s$.
}. This effect will then leak into the tensor perturbation $h_{ab}$ because of the conformal coupling. In Section \ref{sec:scalar_tensor_waveforms}, we demonstrate this for the case of a dark matter particle falling straight down into a black hole. In order to do so, we then first derive the odd- and even-parity master equations for the perturbations of the black hole.

\subsection{Odd-parity sector}
\label{subsec:odd_parity}

To obtain the Regge-Wheeler master equation for the odd-parity perturbations, we decompose the metric perturbation in the Regge-Wheeler gauge as \cite{regge_wheeler_classic}
\begin{eqnarray}
h_{tA} &=& h_0 (r) \epsilon_A^{\ B} \partial_B Y_{lm} \left( \theta, \alpha \right) e^{-i \omega t} \\
h_{rA} &=& h_1 (r) \epsilon_A^{\ B} \partial_B Y_{lm} \left( \theta, \alpha \right) e^{-i \omega t} \\
h_{AB} &=& 0 
\end{eqnarray}
where $Y_{lm}$ are the spherical harmonics, $A = \left( \theta , \alpha \right)$, $\theta$ and $\alpha$ are the polar and azimuthal coordinates on the two-sphere, $\epsilon_2^{\ 2} = \epsilon_3^{\ 3} = 0$, $\epsilon_2^{\ 3} = -1 / \sin \theta$, and $\epsilon_3^{\ 2} = \sin \theta$, and the sum over the multipoles $\left( l, m \right)$ and frequency $\omega$ is implicit. Similarly, we decompose the odd-parity terms of the dark matter's SET as
\begin{eqnarray}
\delta \tilde{T}_{tA} &=& t_0 (r) \epsilon_A^{\ B} \partial_B Y_{lm} \left( \theta, \alpha \right) e^{-i \omega t} \\
\delta \tilde{T}_{rA} &=& t_1 (r) \epsilon_A^{\ B} \partial_B Y_{lm} \left( \theta, \alpha \right) e^{-i \omega t} \\
\delta \tilde{T}_{AB} &=& t_2 (r) \epsilon_{(A}^{\ \ C} \nabla_{B)} \nabla_C Y_{lm} \left( \theta, \alpha \right) e^{-i \omega t} .
\end{eqnarray}
The scalar field $\psi$ does not have an odd-parity component. 

Substituting the above odd-parity decompositions into Eq. \eqref{eq:h_eq}, eliminating $h_0$ using the $\theta\alpha$-component, defining the Regge-Wheeler master function
\begin{equation}
W \left( r \right) = \dfrac{B\left(r\right) h_1 \left(r\right)}{r}, 
\end{equation}
and solving for the $r\alpha$-component, then we obtain Regge-Wheeler equation
\begin{equation}
\label{eq:rw_eq}
\partial^2_{r_*} W + \left( \omega^2 - V_{\text{RW}} \right) W = s_{\text{RW}}
\end{equation}
where
\begin{equation}
\label{eq:rw_potential}
V_{\text{RW}} \left( r \right) = B\left(r\right) \left( \dfrac{l \left(l+1\right)}{r^2} - \dfrac{3 B'\left(r\right)}{r} - 2 \Lambda \right) 
\end{equation}
and
\begin{equation}
\label{eq:rw_source}
s_{\text{RW}} \left( r \right) = - \dfrac{B}{2r^2 F} \left( 2 B \left( r t_1 - t_2 \right) + r \left( t_2 B' + \partial_{r_*} t_2 \right) \right) .
\end{equation}
The coordinate $r_*$ is the usual tortoise coordinate defined by
\begin{equation}
\label{eq:tortoise}
r_* ' \left( r \right) = 1 / B\left( r \right) .
\end{equation}
The primes appearing in Eqs. \eqref{eq:rw_potential}, \eqref{eq:rw_source}, and \eqref{eq:tortoise} are derivatives with respect to $r$.

\subsection{Even-parity sector}
\label{subsec:even_parity}

To describe the perturbations of the even-parity sector, we decompose the metric perturbation in the Regge-Wheeler gauge as \cite{regge_wheeler_classic, zerilli_classic}
\begin{eqnarray}
h_{tt} &=& B(r) H_0 \left( r \right) Y_{lm} \left(\theta, \alpha\right) e^{-i \omega t} \\
h_{tr} &=& H_1 \left( r \right) Y_{lm} \left(\theta, \alpha \right) e^{-i \omega t} \\
h_{rr} &=& H_2 \left( r \right) Y_{lm} \left(\theta, \alpha \right) e^{-i \omega t} / B(r) \\
h_{tA} &=& 0 \\
h_{rA} &=& 0 \\
h_{AB} &=& r^2 K \left( r \right) \gamma_{AB} Y_{lm} \left( \theta, \alpha \right) e^{-i \omega t}
\end{eqnarray}
and the scalar perturbation as
\begin{equation}
\label{eq:psi_decomp}
\psi = \dfrac{\chi \left( r \right)}{r} Y_{lm} \left( \theta, \alpha \right) e^{- i \omega t} .
\end{equation}
Similarly, we decompose the source terms appearing in Eqs. \eqref{eq:h_eq} and \eqref{eq:psi_eq} as
\begin{eqnarray}
\label{eq:delta_T_tt} \delta \tilde{T}_{tt} &=& B(r) T_0(r) Y_{lm} \left( \theta, \alpha \right) e^{-i \omega t} \\
\label{eq:delta_T_tr} \delta \tilde{T}_{tr} &=& T_1(r) Y_{lm} \left( \theta, \alpha \right) e^{-i \omega t} \\
\label{eq:delta_T_rr} \delta \tilde{T}_{rr} &=& T_2(r) Y_{lm} \left( \theta, \alpha \right) e^{-i \omega t} / B(r) \\
\label{eq:delta_T_tA} \delta \tilde{T}_{tA} &=& t_0 (r) \partial_A Y_{lm} \left( \theta, \alpha \right) e^{-i \omega t} \\
\label{eq:delta_T_rA} \delta \tilde{T}_{rA} &=& t_1 (r) \partial_A Y_{lm} \left( \theta, \alpha \right) e^{-i \omega t} \\
\label{eq:delta_T_AB} \delta \tilde{T}_{AB} &=& r^2 \bigg( T_k \left( r \right) \gamma_{AB} Y_{lm} \left( \theta, \alpha \right) \\
& & \phantom{ggg} + T_g \left( r \right) \nabla_A \nabla_B Y_{lm} \left( \theta, \alpha \right) \bigg) e^{-i \omega t} \nonumber
\end{eqnarray}
and
\begin{equation}
\label{eq:delta_rho_decomp}
\delta \tilde{\rho} = \delta m Y_{lm} e^{-i \omega t} .
\end{equation}

By substituting the above expressions into Eq. \eqref{eq:h_eq}, we will be able to eliminate $H_2$ by using the $\theta\alpha$-component. The $t\theta$-, $tr$-, and $r\theta$-components can then be used to obtain $\left( H_0', H_1', K' \right)$ and together with the $rr$-component obtain an algebraic constraint on $H_0$, $H_1$, and $K$. This leaves two remaining perturbations, $H_1 = \omega \mathcal{R}$ and $K$, which we decouple by writing down
\begin{eqnarray}
K \left( r \right) &=& f_1 \left( r \right) \hat{K} \left( r \right)+ f_2\left( r \right) \hat{R} \left( r \right) \\
\mathcal{R} \left( r \right) &=& f_3 \left( r \right) \hat{K} \left( r \right)+ f_4 \left( r \right) \hat{R} \left( r \right) .
\end{eqnarray}
Following the footsteps of Zerilli \cite{zerilli_classic}, demanding that the wave function $\hat{K}$ satisfies a Schr\"{o}dinger-like equation, then we identify
\begin{eqnarray}
f_1 &=& \dfrac{6 M^2+M \left(\Lambda  r^3+3 r \sigma \right)+r^2 \sigma  (\sigma +1)}{r^2 (3 M+r \sigma )} \\
f_2 &=& 1 \\
f_3 &=& \dfrac{i \left(- f_1 r^2+3 M+r \sigma \right)}{r B} \\
f_4 &=& - \dfrac{ i r } {B} 
\end{eqnarray}
where $2 \sigma = l(l+1) - 2$. At this point, the linearized equations for $\left( \hat{K}, \hat{R} \right)$ becomes
\begin{equation}
\label{eq:partial_kh}
\begin{split}
\partial_{r_*} \hat{K} - \hat{R} = &-\dfrac{i B r (r T_1+2 t_0)}{2 F \omega  (3 M+r \sigma )}\\
&+\frac{F' \left(\chi  (3 M-r)-r^2 \partial_{r_*}\chi\right)}{F r (3 M+r \sigma )}
\end{split}
\end{equation}
and
\begin{widetext}
\begin{equation}
\label{eq:partial_rh}
\begin{split}
\partial_{r_*} \hat{R} - \left( \omega^2 - V_{\text{Z}} \right) \hat{K} = 
& \dfrac{i B \left(r^3 \sigma  B T_1+2 t_0 \left(6 M^2+M \left(\Lambda  r^3+3 r \sigma \right)+r^2 \sigma  (\sigma +1)\right)\right)}{2 r \omega  F (3 M+r \sigma )^2} \\
& + \frac{r B (2 B t_1-2 (3 M+r \sigma ) T_g+r T_2)}{2 F (3 M+r \sigma )} \\
& + \chi \left(\frac{B F' \left(9 M^2+6 M r \sigma +r^2 \sigma  (2 \sigma +1)\right)}{r^2 F (3 M+r \sigma )^2}-\frac{r \omega ^2 F' }{F\left(3 M+r \sigma\right) }\right) \\
& + \frac{F' \left(-3 M^2+\Lambda  M r^3-3 M r \sigma +r^2 \sigma \right) \partial_{r_*} \chi }{r F (3 M+r \sigma )^2} .
\end{split}
\end{equation}
\end{widetext}
By differentiating Eq. \eqref{eq:partial_kh} with respect to $r_*$ and eliminating $\partial_{r_*} \hat{R}$ using Eq. \eqref{eq:partial_rh}, then we finally obtain the Zerilli master equation:
\begin{equation}
\label{eq:zerilli_eq}
\partial^2_{r_*} \hat{K} + \left( \omega^2 - V_{\text{Z}} \right) \hat{K} = s_{\text{Z}} + s_{C}
\end{equation}
where the potential, $V_{\text{Z}}$, and source terms, $s_{\text{Z}}$ and $s_C$, are given by
\begin{widetext}
\begin{equation}
V_{\text{Z}} = \dfrac{2 r B \left(9 M^3+M^2 \left(9 r \sigma -3 \Lambda  r^3\right)+3 M r^2 \sigma ^2+r^3 \sigma ^2 (\sigma +1)\right)}{ r^4 (3 M+r \sigma )^2} ,
\end{equation}
\begin{equation}
\label{eq:sZ}
\begin{split}
18 r^2 \omega  F (3 M+r \sigma )^2 s_{\text{Z}} = 3i r B \bigg[ & 2 t_0 \left(18 M^2+3 M r \left(4 \Lambda  r^2+\sigma -3\right)+r^2 \sigma  \left(2 \Lambda  r^2+3 \sigma +3\right)\right) \\
& +r \bigg (2 T_1 \left(9 M^2+3 M r \left(2 \Lambda  r^2-\sigma -3\right)+\Lambda  r^4 \sigma \right) \\
& +2 i \omega  \left(6 M+\Lambda  r^3-3 r\right) (3 M+r \sigma ) t_1 \\
& +3 i r (3 M+r \sigma ) \left(2 \omega  (3 M+r \sigma ) T_g+2 i \partial_{r_*} t_0 + i r \partial_{r_*} T_1-r \omega T_2 \right) \bigg) \bigg] ,
\end{split}
\end{equation}
and 
\begin{equation}
\label{eq:sC}
s_C = -\dfrac{F' \left(\chi \left(r \omega ^2 (3 M+r \sigma )-2 \sigma  (\sigma +1) B\right)+6 M B \partial_{r_*} \chi + r (3 M+r \sigma ) \partial^2_{r_*} \chi \right)}{F (3 M+r \sigma )^2} .
\end{equation}
\end{widetext}
The source $s_{\text{Z}}$ comes directly from the dark matter perturbation $\delta \tilde{T}_{ab}$. On the other hand, the source term $s_C \sim F' \chi$ comes from the conformal coupling between the Ricci scalar and the scalar field in the action
\footnote{The expression given by Eq. \eqref{eq:sC} agrees with Eq. (9) of Ref. \cite{bh_spectro_tattersall} in the Schwarzschild limit.
}.
The direct effect of any dark sector interaction on $\chi$ will therefore enter the Zerilli equation through $s_C$.

The response of the scalar field on the dark interactions is of course encoded in its own wave equation. By substituting the even-parity decomposition of the scalar perturbation and the sources to the scalar wave equation (Eq. \eqref{eq:psi_eq}), we easily end up with
\begin{equation}
\label{eq:psi_zerilli_eq}
\partial^2_{r_*} \chi + \left( \omega^2 - V_{\chi} \right) \chi = s_q + s_h
\end{equation}
where
\begin{equation}
V_{\chi} = \dfrac{B \left(6 M+r \left(r^2 \left(3 \mu ^2-2 \Lambda \right)+6 (\sigma +1)\right)\right)}{3 r^3} ,
\end{equation}
\begin{equation}
\label{eq:s_q}
s_q = q r B \delta m ,
\end{equation}
and
\begin{equation}
s_h = s r B \left(-T_0+T_2-2 \left( \sigma +1 \right) T_g+2 T_k \right) .
\end{equation}
We remind that $\delta m$ is the multipole of the density (Eq. \eqref{eq:delta_rho_decomp}) and the $\delta T_i$'s are the multipoles of the stress-energy tensor (Eqs. \eqref{eq:delta_T_tt}, \eqref{eq:delta_T_tr}, \eqref{eq:delta_T_rr}, \eqref{eq:delta_T_tA}, \eqref{eq:delta_T_rA}, and \eqref{eq:delta_T_AB}). The imprint of the source $s_q$, triggered by dark sector interactions (Eq. \eqref{eq:q_def}), on $\chi$, which then sources the metric perturbations $h_{ab} \sim \hat{K}$, is the physical effect of most interest in this paper. The other source term $s_h \sim s$ (Eq. \eqref{eq:s_def}) comes from the conformal coupling of the scalar field with the Ricci scalar.

Later, the effective mass of the scalar perturbation $\mu$ will be set to zero for simplicity in the sense that the leading-order asymptotic solution of the scalar is $\chi \sim e^{\pm i\omega r_*}$ at the infinity and the event horizon. Noted, by setting $\mu = 0$, we are avoiding tachyonic instabilities but as a price are leaving some novel effects due to the effective mass. We refer the reader interested in the influence of nonzero mass $\mu$ to the scalar's quasinormal spectrum to Refs. \cite{st_horndeski_qnm_tattersall, bh_spectro_tattersall, anomalous_qnms_lagos}.

In the next section, we present the results of numerical integration of the Zerilli (Eq. \eqref{eq:zerilli_eq}) and scalar wave equations (Eq. \eqref{eq:psi_zerilli_eq}) for a radially-plunging dark matter particle.

\section{Scalar and tensor waveforms for an orbiting dark matter particle}
\label{sec:scalar_tensor_waveforms}

The Regge-Wheeler, Zerilli, and the scalar wave equations obtained in the previous section can be numerically integrated to obtain waveforms valid up to first-order of black hole perturbation theory. We do so in this section for a radially-plunging dark matter particle in an asymptotically-flat Schwarzschild black hole $\left( \Lambda = 0 \right)$. This setup allows us to focus on the even-parity modes where dark sector interaction effects can be found.

\subsection{Numerical method}
\label{subsec:numerical_method}

We use the geometrized units $c = 8 \pi G = 1$ throughout. This means that a black hole of mass $M = \nu$ corresponds to a mass $\nu \times 8 \pi M_\odot$ Solar masses where $M_\odot \approx 1.9885 \times 10^{30} \ \text{kg}$ \cite{almanac}. On the other hand, a coordinate $r = \gamma$ corresponds to a distance of $\gamma \times M_\odot \approx 37 \gamma \ \text{km}$. Similarly, a time coordinate $\tau$ and frequency $\omega$ imply a time scale of $\tau \times 8 \pi M_\odot/c \approx 3.1113 \ \text{ms}$ and frequency $\omega \times c/ \left( 8\pi M_\odot \right) \approx 321.41 \ \text{Hz}$. The conformal coupling $s$ (Eq. \eqref{eq:s_def}) and interaction charge $q$ (Eq. \eqref{eq:q_def}) are dimensionless quantities. Useful conversion factors to keep in mind are $1 \ \text{s} = 299792 \ \text{km}$, $1 \ M_\odot = 37.113 \ \text{km}$, and $c^4/\left( 8 \pi G \right) = 0.026945 \ M_\odot/\text{km}$.

The stress-energy tensor for a particle falling straight into a black hole is presented in Appendix \ref{sec:set_plunge}. The sources of the wave equations (Eqs. \eqref{eq:zerilli_eq} and \eqref{eq:psi_zerilli_eq}) can be prepared by substituting Eqs. \eqref{eq:set_fp_decomp} and \eqref{eq:delta_m_fp_decomp} into Eqs. \eqref{eq:sZ} and \eqref{eq:sC}. The numerical integration then proceeds after providing the parameters of the black hole, $\left( M , \Lambda \right)$, the geodesic constants, $\left( \varepsilon, z\right)$, the mode label, $\left(l, m\right)$, and the theory constants, $\left( \mu, s, q, F\left(\varphi\right), F'\left(\varphi\right) \right)$ that explicitly appear in the wave equations.

We proceed numerically in the manner briefly reviewed in Appendix \ref{sec:integration}, i.e., first look for the homogeneous solution satisfying the boundary conditions given by Eq. \eqref{eq:z1} and integrate over the source via Eq. \eqref{eq:y_gen_infinity}. We assume the detector will be at infinity where the spacetime is asymptotically flat. The integration is then carried out for equal intervals $\Delta \omega$ in the frequency range $\omega \in \left(-3 , 3\right)$. This is done for both wave equations but with the caveat that the particular scalar solution to Eq. \eqref{eq:psi_zerilli_eq} will also source the tensorial mode solution of Eq. \eqref{eq:zerilli_eq}. The scaling influence of the interaction charge $q$ on the scalar mode then appears in the tensorial mode. 

For this analysis, we consider a choice of parameters that highlights the influence of dark sector interaction on both the scalar and tensor waveforms. In particular, we set $\mu = 0$ to keep the analysis of the scalar mode simple
\footnote{
It should be stressed that more interesting effects, specifically on the quasinormal spectra, can be expected with $\mu \neq 0$ (see Refs. \cite{st_horndeski_qnm_tattersall, bh_spectro_tattersall, anomalous_qnms_lagos}). However, our focus in this paper is the interaction charge which enters the source. We leave the broader analysis for future work.
}.
We then focus on the stronger $\left(l, m\right) = \left(2, 0\right)$ mode. The mass of the black hole is chosen to be $M = 1/2$. The geodesic parameters considered are $\varepsilon = 10$ and $z = r_0/\left(2M\right) = 10$ where $r_0$ is the dark matter particle's position at $t = 0$. The rest of the parameters have been chosen to highlight the effects of the interaction on the waveforms. The nonzero theory constants considered are $s = 10^{-2}$, $F\left(\varphi\right) = 1/2$, and $F'\left(\varphi\right) = 10^{-2}$. Most importantly, to demonstrate the main topic of this paper, we perform the integration with these parameters for $q = 0$, $q = 1$, and $q = -1$.

Looking back at the sources of the coupled scalar-tensor wave equations (Eqs. \eqref{eq:sC} and\eqref{eq:s_q}), it becomes clear that the conformal potential $F\left( \phi \right)$ must be varying; otherwise, the dark interaction would not be observable in the tensor sector. Echoing the discussion in the last paragraph of Sec. \ref{subsec:linearized_equations}, the effect we are looking at is a magnification (or deterioration) of the scalar waveform by a factor $-q/s$. Provided that there is conformal coupling, i.e., $F' \neq 0$, we can expect this to translate to an enhancement of the conformal source term $s_C$ (Eq. \eqref{eq:sC}) in the Zerilli equation by the same factor. The values $s, F' \sim 10^{-2}$ and $q \sim 1$ were found to be sufficient to reveal this effect. Larger (smaller) values than the ones chosen here only make the effect more pronounced (suppressed). All this is transparent in the results.

For the stability of the numerical integration, we include the first five leading asymptotic corrections to the scalar and tensor waveforms at the horizon and at infinity. We keep the working precision to 30 digits for internal evaluation of the solutions of Eqs. \eqref{eq:zerilli_eq} and \eqref{eq:psi_zerilli_eq} satisfying the prescribed boundary conditions. Then, we keep 5 significant digits of accuracy in performing the integration over the sources in order for the integrals to converge in a reasonable time scale of seconds for each $\omega$ in $\left( -3, 3 \right)$ in a standard personal machine (with a processor Intel Core I7). The ``double exponential" option of Mathematica's \textit{NIntegrate} which specializes on the evaluation of highly oscillatory integrals was implemented. All of the tiny details of the numerical scheme are transparently communicated in the Mathematica notebook \textit{integration.nb} publicly available in the \href{https://github.com/reggiebernardo/notebooks}{author's github repository}. The results are presented in the next section.

\subsection{Results}
\label{subsec:results}

The power spectra for both the scalar and tensor modes will be assumed to be given by $\omega^2 | \Psi \left( \omega \right) |^2$ where $\Psi = \chi, \hat{K}$. This assumes the Isaacson stress-energy tensor for the gravitational wave energy and neglects interference effects between scalar and tensor modes which generally occur when $F'\left( \varphi \right) \neq 0$ \cite{set_saffer}. The interference can be expected to be proportional to $F'$ and the choice $F' \sim 10^{-2}$ in the simulations should keep this subdominant by the same factor of $10^{-2}$ compared to the individual scalar and tensor contributions to the gravitational wave energy. Nonetheless, this simple assumption allows us to look at the individual scalar and tensor power spectra which will backreact to the orbital evolution at higher-order in perturbation theory. The power spectra including corrections to the Isaacson stress-energy tensor for GWs will be considered as a future work.

\begin{figure}[h!]
\center
	\subfigure[ Scalar power spectrum ]{
		\includegraphics[width = 0.45 \textwidth]{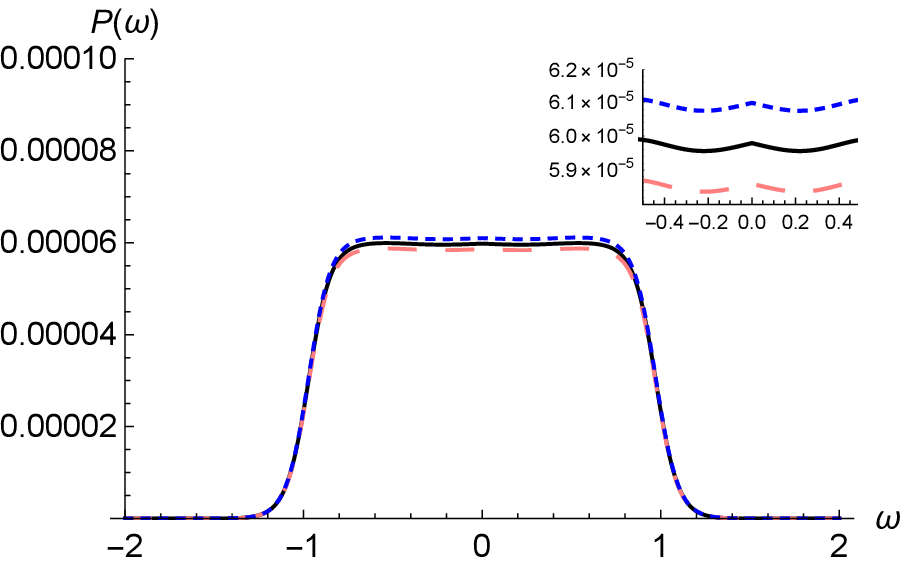}
		}
	\subfigure[ Tensor power spectrum ]{
		\includegraphics[width = 0.45 \textwidth]{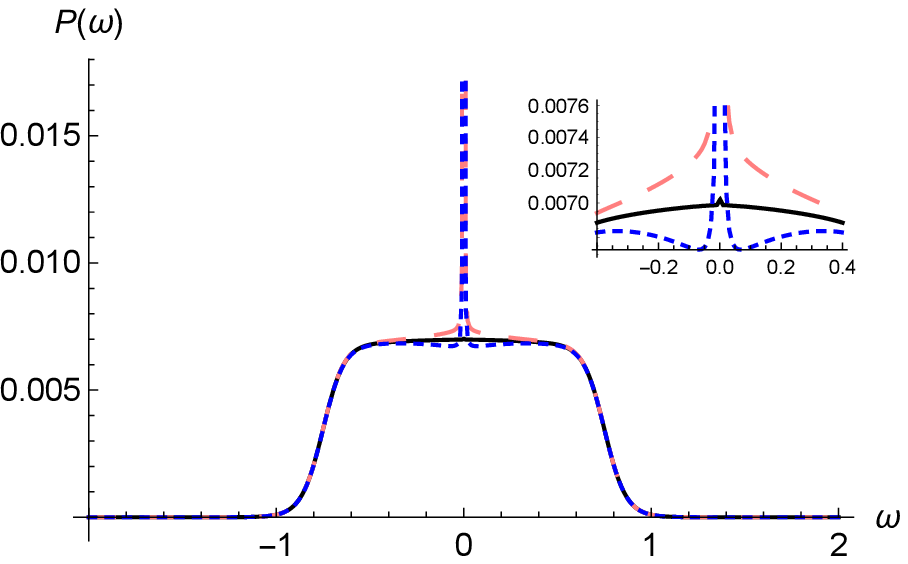}
		}
\caption{\textit{Power spectra} of the scalar and tensor modes. (a) Scalar power spectra for $q = 0$ multipled by $10^{4}$ (black solid line), $q = 1$ (pink long-dashed line), and $q = -1$ (blue short-dashed line). (b) Tensor power spectra for $q = 0$ (black solid line), $q = 1$ (pink long-dashed line), and $q = -1$ (blue short-dashed line). Inset shows a magnified view of logarithmic power spectra in the region $\omega \in \left(-0.4, 0.4\right)$.}
\label{fig:ps}
\end{figure}

The individual power spectra of the scalar and tensor modes are shown in Fig. \ref{fig:ps}. This shows the characteristic shape, e.g., a low-frequency plateau being trimmed off exponentially at the quasinormal frequency, of the energy spectrum per mode of a Schwarzschild black hole with a radially-falling particle \cite{gw_power_spectrum_cardoso_1, gw_power_spectrum_cardoso_2}. The exponential decay indicative of the quasinormal modes' taking over is also reflected in the ringdown stages of the time-domain waveforms.

The scaling effect of the interaction on the scalar modes can be seen in action in Fig. \ref{fig:ps}(a) where the power spectrum for the noninteracting case $q = 0$ have been multipled by $10^4$ in order to accommodate the $10^2$ factor difference in $\chi$ when $q$ is $O\left(1\right)$. Also, because of the small but nonzero $s$, there is a resolvable difference between the $q = 1$ and $q = -1$ scalar power spectra (shown in the inset of Fig. \ref{fig:ps}(a)). The tensor power spectra shown in Fig. \ref{fig:ps}(b) also shows that the more dramatic effect of dark sector interactions can be found in the low-frequency limit. Specifically, Fig. \ref{fig:ps}(b) shows that scalar and tensor modes for $q =1$ have interfered constructively at low-$\omega$ and so increased the power spectrum compared with the noninteracting case $q = 0$. On the other hand, the scalar and tensor modes for $q = -1$ appear to have intefered destructively. These results highlight the places, in this case, low frequencies, where the effects of the dark sector interaction can be expected to be significant.

\begin{figure}[h!]
\center
	\subfigure[ Scalar waveform ]{
		\includegraphics[width = 0.45 \textwidth]{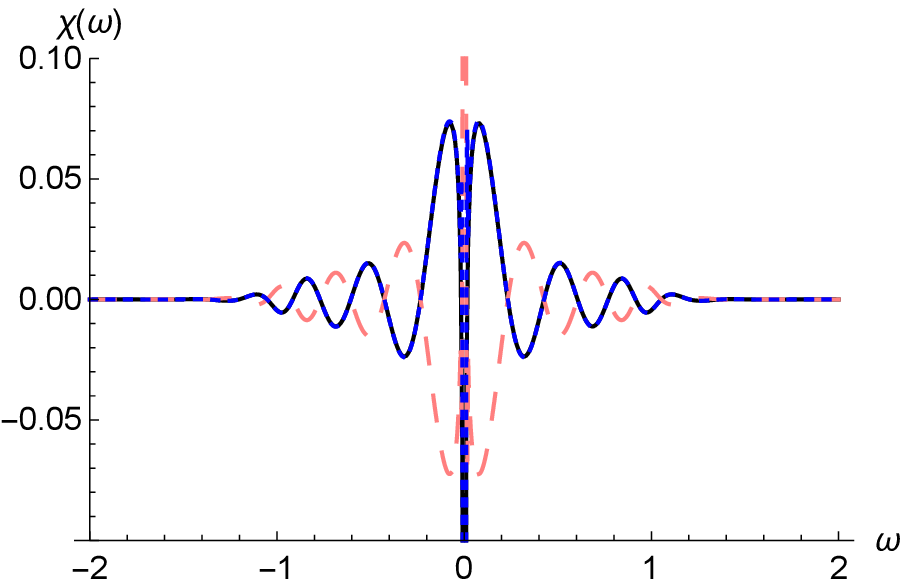}
		}
	\subfigure[ Tensor waveform ]{
		\includegraphics[width = 0.45 \textwidth]{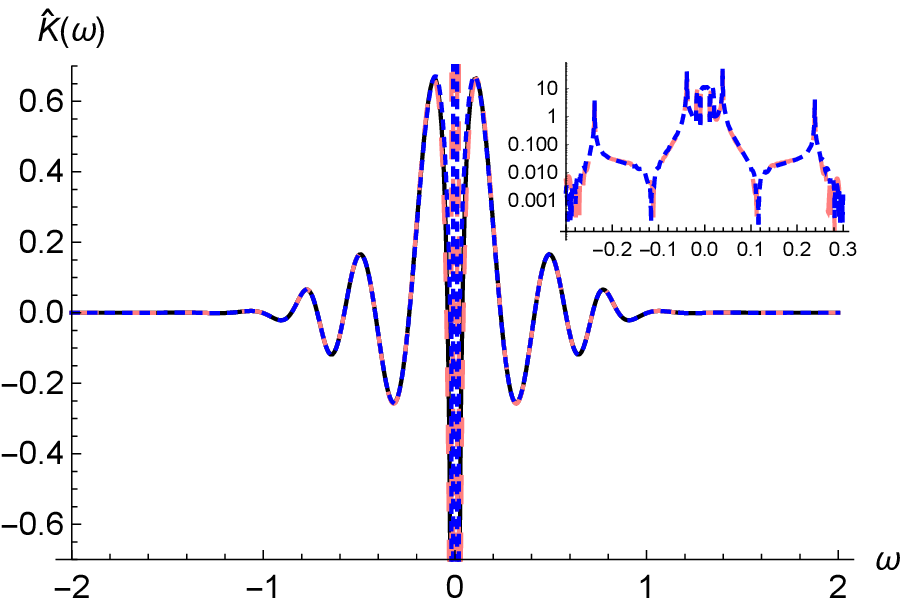}
		}
\caption{Real part of the \textit{frequency-domain waveforms}. (a) Scalar waveforms for $q = 0$ rescaled by $10^{2}$ (black solid line), $q = 1$ (pink long-dashed line), and $q = -1$ (blue short-dashed line). (b) Tensor waveforms for $q = 0$ (black solid line), $q = 1$ (pink long-dashed line), and $q = -1$ (blue short-dashed line). Inset shows the logarithmic resolvable error, $\ln \left( | \Delta \hat{K}/\hat{K} | \right)$, in the region $\omega \in \left(-0.4, 0.4\right)$ when the $q = 1$ (pink long-dashed) and $q =-1$ (blue short-dashed) waveforms are compared with the noninteracting case $q = 0$.}
\label{fig:fd_waveforms}
\end{figure}

Fig. \ref{fig:fd_waveforms} shows the frequency-domain waveforms for both the scalar and tensor modes.
The scaling effect of the interaction is again in full display in Fig. \ref{fig:fd_waveforms}(a) where the waveform for $q = 0$ was multipled by a factor of $10^2$. The reflected orientation of the $q = 1$ waveform is also in action. The corresponding tensor waveforms are shown in Fig. \ref{fig:fd_waveforms}(b). The inset shows the logarithmic resolvable error, $ \ln \left( | \Delta \hat{K}/\hat{K} | \right)$, when the $q = \pm 1$ waveforms are compared to the noninteracting case $q = 0$. This shows indeed that the significant contamination of the tensor waveform by the dark sector interaction occurs in the low-frequency limit. This also supports the power spectrum results presented previously.

\begin{figure}[h!]
\center
	\subfigure[ Scalar waveform ]{
		\includegraphics[width = 0.45 \textwidth]{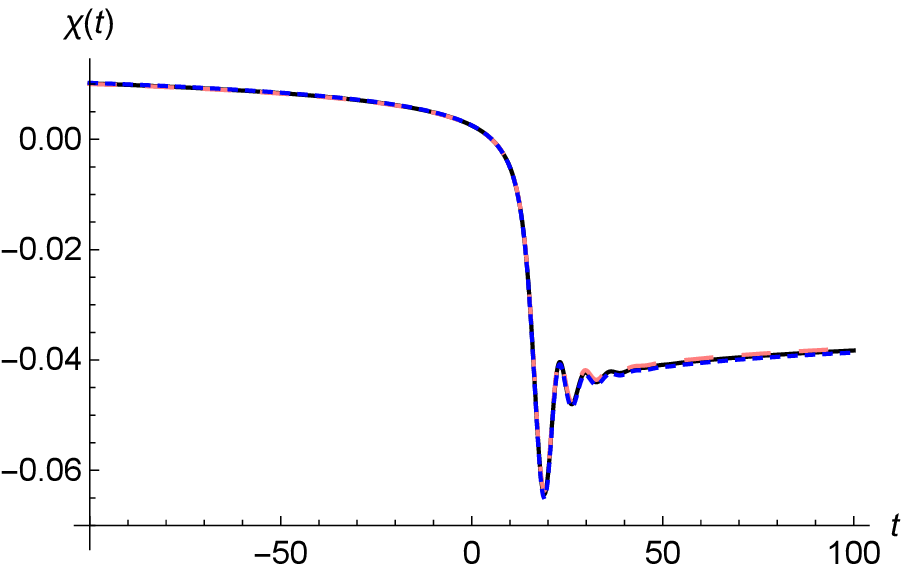}
		}
	\subfigure[ Tensor waveform ]{
		\includegraphics[width = 0.45 \textwidth]{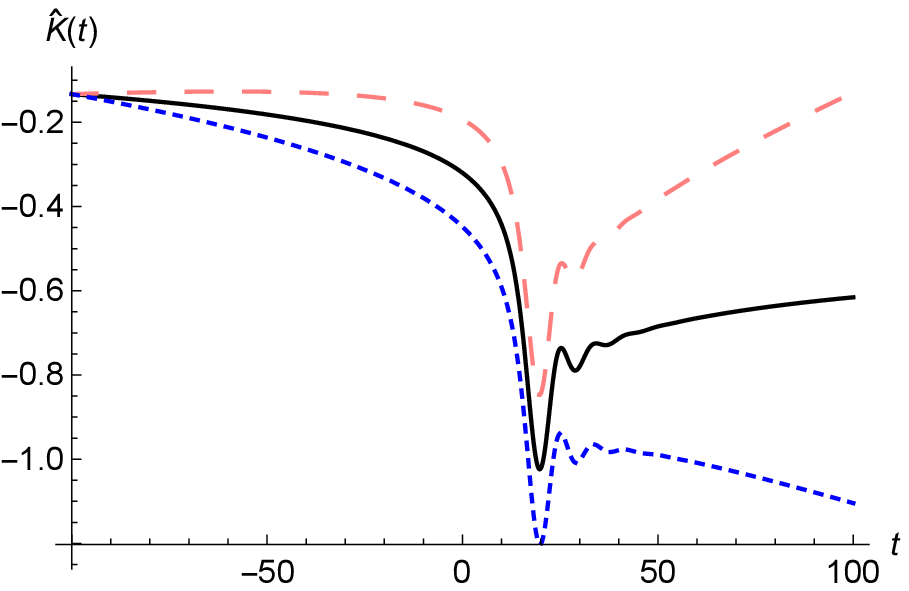}
		}
\caption{Real part of the \textit{time-domain waveforms}. (a) Scalar waveforms for $q = 0$ rescaled by a factor $10^{2}$ (black solid line), $q = 1$ reflected, i.e., $-\chi$ is shown, (pink long-dashed line), and $q = -1$ (blue short-dashed line). (b) Tensor waveforms for $q = 0$ (black solid line), $q = 1$ (pink long-dashed line), and $q = -1$ (blue short-dashed line). The $q = -1$ tensor waveform is reflected and the resulting waveforms for $q = \pm 1$ have been vertically adjusted to match the $q = 0$ waveform at $t = -100$. The adjustments were made to highlight the differences in the shape of the time-domain waveforms due to the dark sector interaction.}
\label{fig:td_waveforms}
\end{figure}

The corresponding time-domain waveforms are presented in Fig. \ref{fig:td_waveforms}. This displays characteristic waveforms for a particle falling straight into a black hole \cite{gw_power_spectrum_cardoso_1, gw_power_spectrum_cardoso_2}. The late-time ringdown is reflective of the exponential damping observed earlier in the power spectrum.

In Fig. \ref{fig:td_waveforms}(a), the scalar waveform for $q = 0$ was amplified by a factor $10^2$ and the one for $q = 1$ was reflected, i.e., $\chi \rightarrow -\chi$. This once again supports the scaling effect of the dark sector interaction on the scalar modes that was discussed in the last paragraph of Section \ref{subsec:linearized_equations}. The low-$\omega$ contamination of the tensor waveform due to dark sector interaction is shown in Fig. \ref{fig:td_waveforms}(b). In this particular plot, the waveforms have been adjusted to agree at early-times so that the overall effect on the shape of the waveform will be highlighted. This then shows that dark sector interactions affect the overall orientation of the time-domain tensor waveforms in a distinct way.

Indeed, the radially-plunging setup is an astrophysically improbable scenario. Nonetheless, we can view the concrete numerical results of this section as conclusive that dark sector interactions can directly imprint on GWs. This supports the assertion that general relativistic black holes can be used as a direct probe of the existence of dark sector interactions. The contributions from the higher-order modes ($l \geq 3$) can be confirmed to be subdominant compared to the $l = 2$ mode. However, in any practical orbit, more than a single mode matters and all these should be calculated in order to get a precise result. This will be considered in a future work.

\section{Conclusions}
\label{sec:conclusions}

We have shown that the existence of dark sector interactions can be inferred from the waveforms generated by general relativistic black hole perturbations. This result is encapsulated by Eqs. \eqref{eq:h_eq}, \eqref{eq:psi_eq}, and \eqref{eq:cdm_conservation}, described in Section \ref{sec:bh_perturbations}, and fleshed out in Section \ref{sec:scalar_tensor_waveforms}. It should be stressed that the fluid variables appearing in these equations correspond to the total dark matter field, including the effects of dark sector interactions, and that these are equations in the physical Jordan frame. The scalar field's effective mass $\mu$ (Eq. \eqref{eq:mu_def}) also generalizes the corresponding expression presented in Ref. \cite{st_horndeski_qnm_tattersall} to nonzero $\Lambda$. Most importantly, the interaction charge $q$ (Eq. \eqref{eq:q_def}), the residue of dark sector interactions on a general relativistic black hole, is introduced for the first time. This interaction charge directly scales the scalar waveform and can be expected to influence the tensor waveform as long as there is any conformal coupling between the scalar field and the Ricci scalar in the action. We have obtained the Regge-Wheeler equation (Eq. \eqref{eq:rw_eq}) and the coupled Zerilli and scalar wave equations (Eqs. \eqref{eq:zerilli_eq} and \eqref{eq:psi_zerilli_eq}) in a Schwarzschild-(anti) de Sitter black hole and have explicitly demonstrated the effect of dark sector interactions on the scalar and tensor waveforms generated by a dark matter particle falling straight down into a Schwarzschild black hole.

Constraining dark sector interactions must be sought further. One way forward is to compute the ratio $q/s$ using Eqs. \eqref{eq:s_def} and \eqref{eq:q_def} and inherit the constraints coming from cosmological observations. For example, in coupled quintessence \cite{coupled_quint_pettorino, ide_gomez}, the factor $\beta = I_{1\phi} / \left( 1 + I_1 \right)$ was determined to be of order $\beta \sim 10^{-2}$ using various cosmological data sets. On the other hand, an estimate of the conformal factor $F^{\prime 2}/F$ based on the results of Ref. \cite{horndeski_constraint_komatsu2019} leads to $F^{\prime 2}/F \sim 10^{-3}/\phi_0^{\prime 2}$ where $\phi'_0 = d \phi/d a |_{a = 1}$ and $\phi(a)$ is the dark energy field. Putting these together leads to $q / s \sim 10/\phi_0^{\prime 2}$. This crude, order of magnitude estimate supports the potential observability of the waveform signatures first described in this paper. However, a more rigorous approach to constraining the dark sector parameters $q$ and $s$ in the black hole setting should be pursued. We refer the reader to Refs. \cite{st_horndeski_qnm_tattersall, bh_spectro_tattersall, anomalous_qnms_lagos} on constraining the effective mass $\mu$ through quasinormal modes.

It is worth remarking again that the general relativistic black hole setting will fail to probe dark sector interactions of the form given by Eq. \eqref{eq:Lc2}. This inherent limitation prompts a follow-up question: ``Can the potential $I_2$ be probed using gravitational waves generated in a different setting?" This will be considered in a different paper.

This paper leaves some more analysis for future work. The most obvious one being that astrophysical orbits are far richer than just the case of a dark matter particle falling straight into a compact object and involve more than a few modes in spherical harmonics. The analysis of the waveform and the influence of dark sector interactions on quasicircular orbits will be discussed in a different paper. Extensions to the perturbations of rotating black holes and even possibly to relativistic stars can also be done. The important question of the cosmological significance of general relativistic black holes in scalar-tensor theories should also be taken up in a future work. A constant scalar field will not be able to drive cosmic acceleration and implies that a kind of screening mechanism must exist that reduces the scalar to a constant on black holes. Lastly, it is interesting to see whether the results obtained in this paper will hold outside of the framework of Horndeski theory and the two interaction Lagrangians considered in this paper.

\section*{Acknowledgements}
The author is grateful to Che-Yu Chen for constructive feedback on an earlier version of the manuscript. The author acknowledges the use of the `xAct' package \cite{xact} and its derivatives `xPert' \cite{xpert} and `xCoba' \cite{xcoba}.

\appendix

\section{Stress-energy tensor for radially-plunging particle}
\label{sec:set_plunge}

We present the stress-energy tensor (SET) for a radially-plunging particle. 

The SET of a test particle satisfying the geodesic equation can be written as
\begin{equation}
\label{eq:set_particle}
T_{ab} = \frac{u_a u_b}{u^t} \rho^{(3)} \left( t, \vec{x} \left( t \right) \right)
\end{equation}
where $u^a$ is the particle's four-velocity and $\rho^{(3)} \left( t, \vec{x} \left( t \right) \right)$ is the three-density of the particle on the trajectory $\vec{x}\left(t\right)$. The four-velocity components of the particle are given by
\begin{eqnarray}
u^t &=& \frac{ \varepsilon }{B \left(r\right)} \\
u^r &=& - \sqrt{ \varepsilon^2 - B \left(r\right) } \\
u^\theta &=& 0 \\
u^\varphi &=& 0
\end{eqnarray}
and the particle's coordinate velocity is given by
\begin{equation}
\frac{dr_p}{dt_p} = - B\left( r_p \right) \sqrt{ 1 - \frac{ B\left( r_p \right) }{ \varepsilon^2 } } 
\end{equation}
where $\varepsilon$ is the conserved energy on the geodesic and $B\left(r\right)$ is the metric function (Eq. \eqref{eq:black_hole_metric}). There are no odd-parity components. The SET can be written in matrix form:
\begin{equation}
T_{ab} = 
\frac{B\left(r\right)}{\varepsilon} \rho^{(3)} \left( t, \vec{x} \left( t \right) \right) u_a u_b 
\end{equation}
where
\begin{equation}
\begin{split}
\dfrac{\rho^{(3)} \left( t, \vec{x} \left( t \right) \right)}{m} =  \dfrac{ \delta \left( r - r_p \left( t \right) \right) }{r^2} \delta \left( \cos \theta- \cos \theta_0 \right) \delta \left( \alpha - \alpha_0 \right)
\end{split}
\end{equation}
and
\begin{equation}
u_a u_b = \left( 
\begin{array}{c c c c}
\varepsilon^2 & \ \ \dfrac{\varepsilon \sqrt{\varepsilon^2 - B}}{B} \ \ \ \ & \ \ \ \ 0 \ \ \ \  & \ \ \ \ 0 \ \ \ \ \phantom{\frac{\frac{1}{1}}{\frac{1}{1}}} \\
\ \  \dfrac{\varepsilon \sqrt{\varepsilon^2 - B}}{B} \ \  & \ \ \dfrac{\varepsilon^2 - B}{B^2} \ \ \ \  & \ \ \ \ 0 \ \ \ \  & \ \ \ \ 0 \ \ \ \ \phantom{\frac{\frac{1}{1}}{\frac{1}{1}}} \\
\ \ 0 \ \  & \ \ 0 \ \ \ \  & \ \ \ \ 0 \ \ \ \  & \ \ \ \ 0 \ \ \ \ \phantom{\frac{\frac{1}{1}}{\frac{1}{1}}} \\
\ \ 0 \ \  & \ \ 0 \ \ \ \  & \ \ \ \ 0 \ \ \ \ & \ \ \ \ 0 \ \ \ \ \phantom{\frac{\frac{1}{1}}{\frac{1}{1}}}
\end{array} 
\right) .
\end{equation}
Clearly, integrating $\rho^{(3)} \left( t, \vec{x} \left( t \right) \right)$ over all space gives the mass $m$ of the particle. It is easy to show that
\begin{equation}
\delta \left( r - r_p(t) \right) = \dfrac{1}{2\pi} \int_{-\infty}^{\infty} d\omega \frac{ e^{-i\omega \left( t - t_p \left( r \right) \right)} }{ | dr_p/dt | }
\end{equation}
where $t_p \left (r \right)$ is defined by
\begin{equation}
- dt_p \left( r \right) = \dfrac{dr_*(r)}{ \sqrt{1 - \dfrac{B\left(r\right)}{\varepsilon^2} }  }
\end{equation}
and $dr_*(r) = dr / B(r)$ defines the tortoise coordinate $r_*\left(r\right)$. Note that the geodesic energy $\varepsilon \in \left( \sqrt{B}, \infty \right)$ where the limit $\varepsilon = \sqrt{B \left(r_0\right)}$ corresponds to the particle falling from rest at $r = r_0$. In the frequency-domain $\omega$, the particle's SET becomes
\begin{equation}
T_{ab} = \frac{m}{2\pi \varepsilon} \dfrac{ e^{i \omega t_p \left(r\right)} }{r^2 \sqrt{ 1 - \dfrac{B}{\varepsilon^2} } } \delta^{(2)}\left(\Omega - \Omega_0\right) u_a u_b
\end{equation}
where
\begin{equation}
\delta^{(2)}\left(\Omega - \Omega_0\right) = \delta \left( \cos \theta - \cos \theta_0 \right) \delta \left( \alpha - \alpha_0 \right) .
\end{equation}
Using the spherical harmonics closure relation
\begin{equation}
\delta^{(2)}\left(\Omega - \Omega_0\right) = \sum_{l=0}^\infty \sum_{m=-l}^{m=l} Y^*_{lm} \left( \theta_0 , \alpha_0 \right) Y_{lm} \left( \theta , \alpha \right) ,
\end{equation}
we can obtain the source terms entering the Zerilli equation to be
\begin{equation}
\label{eq:set_fp_decomp}
\begin{split}
T_0\left( r \right) &= \dfrac{ T_1 \left( r \right) }{\sqrt{ 1 - \dfrac{B \left( r \right)}{\varepsilon^2} }} = \frac{T_2 \left(r\right)}{ 1 - \dfrac{B \left( r \right)}{\varepsilon^2}  } \\
& = \frac{m \varepsilon}{2\pi} \frac{ e^{i \omega t_p \left(r\right)} }{r^2 B\left( r \right) \sqrt{ 1 - \dfrac{B \left( r \right)}{\varepsilon^2} } } Y^*_{lm} \left( \theta_0, \alpha_0 \right) .
\end{split}
\end{equation}
We also remind the decomposition $\delta \rho = \rho^{(3)}/u^t = \delta m Y_{lm}$ of the source of the scalar mode equation. It can be shown that
\begin{equation}
\label{eq:delta_m_fp_decomp}
\delta m \left( r \right) = \frac{ m }{2\pi \varepsilon} \dfrac{ e^{i \omega t_p \left(r\right)} }{r^2 \sqrt{ 1 - \dfrac{B}{\varepsilon^2} } } Y^*_{lm} \left( \theta_0, \alpha_0 \right) .
\end{equation}

\section{Numerical integration of the wave equation}
\label{sec:integration}

We review the numerical integration of the wave equation. Consider the ordinary differential equation
\begin{equation}
\label{eq:ode_generic}
B^2(r) y''(r) + B(r) B'(r) y'(r) + Q(r) y(r) = S(r)
\end{equation}
where $B$, $Q$, and $S$ are arbitrary functions. For Eqs. \eqref{eq:zerilli_eq} and \eqref{eq:psi_zerilli_eq}, $y = \left( \chi, \hat{K} \right)$, $\partial_{r_*}^2 y = B^2 y'' + B B' y'$, and $Q = \omega^2 - V$. In what follows, we omit the explicit $r$ dependence for brevity. For the asymptotically flat, nonrotating, Schwarzschild black hole, we integrate Eq. (\ref{eq:ode_generic}) for two independent solutions $Z_1$ and $Z_2$ with the following asymptotic behavior:
\begin{equation}
\label{eq:z1}
Z_1 \sim
\begin{cases}
e^{- i\omega r_*} &, \ \ \ \ r \rightarrow r_H \\
\mathcal{I} e^{- i\omega r_*} + \mathcal{R} e^{i\omega r_*} &, \ \ \ \ r \rightarrow \infty 
\end{cases}
\end{equation}
\begin{equation}
Z_2 \sim
\begin{cases}
\mathcal{M} e^{i \omega r_*} + \mathcal{T} e^{-i\omega r_*} &, \ \ \ \ r \rightarrow r_H \\
e^{i\omega r_*} &, \ \ \ \ r \rightarrow \infty . 
\end{cases}
\end{equation}
The Wronskian of these two solutions is given by 
\begin{equation}
W(Z_1,Z_2) = Z_1' Z_2 - Z_1 Z_2' = - \frac{2 i \omega \mathcal{I}}{B} .
\end{equation}
This is to be expected given that $\left( B W \right)' = 0$ follows from Eq. \eqref{eq:ode_generic}. Thus, $B W$ is a constant equal to $-2 i \omega \mathcal{I}$ for all $r$. After obtaining $Z_1$ and $Z_2$, we can write down the general solution
\begin{equation}
\begin{split}
y (r) = \frac{i}{2\omega \mathcal{I}} \bigg( & \phantom{+} Z_1(r) \int^\infty_r dr' \dfrac{ Z_2 (r') S(r') }{ B(r') } \\
& + Z_2 (r) \int_{r_H}^r dr' \dfrac{ Z_1 (r') S(r') }{ B(r') } \phantom{+} \bigg) .
\end{split}
\end{equation}
Setting up the detector at infinity, the waveform should then be
\begin{equation}
\label{eq:y_gen_infinity}
y \left( r \rightarrow \infty \right) \sim \dfrac{i e^{i\omega r_*}}{2\omega \mathcal{I}} \int^\infty_{r_H} dr' \dfrac{ Z_1 (r') S(r') }{ B(r') }  .
\end{equation}

%\bibliographystyle{apsrev4-2}
%\bibliography{bibfile}


\end{document}